\documentclass[aps,floatfix,
	pra,
	showpacs,
  twocolumn,groupedaddress%,superscriptaddress
	]{revtex4}
\usepackage{graphicx}%
\usepackage{dcolumn}% Align table columns on decimal
\usepackage{amssymb,amsmath}
\usepackage{bm}

\usepackage[usenames]{color}

\begin{document}

\definecolor{darkblue}{rgb}{0,0.1,0.9}
\definecolor{darkred}{rgb}{0.9,0.1,0.}
\newcommand{\markg}{{}}
\newcommand{\marko}{{}}

\title{Long-lifetime Polariton BEC in a  Microcavity with an Embedded Quantum Well and Graphene}

\author{O.~L.~Berman}
\affiliation{Physics Department, New York City College of Technology, CUNY, Brooklyn, NY 11201, USA}
%\affiliation{The Graduate School and University Center, City University of New York, New York, NY 10016, USA}

\author{R.~Ya.~Kezerashvili}
\affiliation{Physics Department, New York City College of Technology, CUNY, Brooklyn, NY 11201, USA}
%\affiliation{The Graduate School and University Center, City University of New York, New York, NY 10016, USA}

\author{G.~V.~Kolmakov}
\affiliation{Physics Department, New York City College of Technology, CUNY, Brooklyn, NY 11201, USA}

\date{\today}

\begin{abstract} 
We study the propagation of a Bose-Einstein condensate (BEC) of long-lifetime exciton polaritons
in a high-quality microcavity with an embedded semiconductor quantum well or a graphene layer
using  the Gross-Pitaevskii equation.  
It is shown that in these heterostructures the BEC of the long-lifetime polaritons
can propagate over the distance up to 0.5 mm. 
The obtained results are consistent with the recent experimental observations for GaAs/AlGaAs microcavity.
{\marko It is demonstrated that the BEC density in a polariton trace in a microcavity with embedded %gapped 
graphene at large distances from the excitation spot is higher for the microcavity with higher dielectric constant.}
It is {\marko also} predicted that the propagation of a polariton BEC in a microcavity with  graphene  
is dynamically tunable by changing the gap energy, that makes it potentially useful for applications 
in integrated optical circuits.
\end{abstract}


\maketitle

\section{Introduction}\label{sec:intro}
\vspace{-0.5cm}
Polaritons are the quantum superposition of photons and 
excitons in semiconductors or graphene \cite{Snoke:09,Berman:10c,Carusotto:13}.
In the past years, formation and spreading of the polariton Bose-Einstein condensate (BEC) in  
quasi-two-dimensional structures attract attention of experimentalists and theoreticians
from the point of view of fundamental physics at nanoscales and  due to potential applications 
of  polaritons in integrated  circuits for optical and quantum 
computing \cite{High:08,Liew:10,Menon:10}. Because the effective mass of the polaritons
is much smaller than the atomic masses, the BEC transition temperature is much 
higher than that for atomic BECs \cite{Dalfovo:99}. Another important features of the polariton BEC dynamics is that 
it is essentially non-equilibrium due to the finite polariton lifetime.
The polariton decay is mainly caused by the leakage of photons from an optical microcavity 
and usually the polariton lifetime is of the order of a few picoseconds, 
depending on  the quality Q-factor of the microcavity \cite{Carusotto:13}. 
It was established that the polariton BEC can move in a planar microcavity  in response of the external 
force or field \cite{Sermage:01,Amo:09}. 
However, in ``traditional'' experiments the characteristic length 
over which the BEC can spread in the microcavity
is limited due to relatively short lifetime  of the polaritons \cite{Amo:09}.

The signatures of the polariton BEC were first observed in GaAs-based
microcavity \cite{Deng:02,Deng:03,Deng:07,Balili:06,Balili:07,Balili:09}
and then in CdTe-based ~\cite{Kasprzak:06} and GaN-based
microcavities \cite{Christopoulos:07}. 
%In the present Paper we consider the polaritons formed by the excitons in GaAs 
%quantum well and graphene embedded in a microcavity. 
There are two advantages for observation of a polariton BEC in such heterostructures 
over a conventional exciton BEC~\cite{Hanamura:77}.
First, polaritons have  an effective mass, which is $\sim 10^{-4}$ times lower than the 
exciton mass, that results in much higher critical temperature of the 
BEC transition for polaritons than that for the exciton BEC at the same particle 
density~\cite{Snoke:02,Littlewood:07,Snoke:09}. 
Second, the polariton BEC is less sensitive, compared to the exciton BEC, 
to the disorder caused by unavoidable crystal defects and impurities. 
It is worth noting that the presence of the defects results in the exciton localization
 in a fluctuating potential in a sample and can destroy an 
exciton BEC \cite{Szymanska:02,Marchetti:04,Marchetti:06,Malpuech:07}.

It was found  \cite{Saba:01,Deng:02,Deng:03,Kasprzak:06,Deng:07,Lagoudakis:08,Utsunomiya:08,Nelsen:09,Balili:09,Roumpos:12} %\cite{Deng:10} 
that for optimal conditions 
of the polariton BEC formation the quality Q-factor of the optical microcavity should be enough high.
The increase of the microcavity Q-factor leads to the increase of the polariton 
densities in the microcavity and, also, to longer polariton lifetime thus presenting more 
favorable conditions for the BEC observations.
Also, enough strong exciton-photon coupling (Rabi splitting) is needed thus,  requiring
large quantum well (QW) exciton oscillator strength.
Enough large  polariton-polariton interaction scattering cross-section is also an important factor, which 
provides efficient thermalization in the polariton system and hence  short enough relaxation time for the
transition of the system to a BEC (see Ref. \cite{Snoke:02} for the discussion).
{\marko Rapid expansion of a spatial coherence and the polariton BEC formation dynamics 
in a GaAs-based microcavity has been recently observed in Ref. \cite{Belykh:13}.}

 In the present Article we consider the dynamics of polaritons formed by the cavity photons and 
the excitons in GaAs a quantum well and graphene embedded in a microcavity.  
Our studies are motivated by experiments \cite{Sermage:01} and \cite{Nelsen:12}
where the directional propagation of exciton polaritons have been observed in the presence of an external force. 
In particular, in Ref.\  \cite{Nelsen:12} the possibility of synthesis of a semiconductor 
microcavity with the Q-factor exceeding $10^5$ has been demonstrated for the first time.
That allowed one to observe  the polaritons with extremely long 
lifetime $\sim 100$ ps, that is $\sim 30\times$ longer than that in the traditional experiments \cite{Carusotto:13}.
It was found \cite{Nelsen:12} that the long-lifetime 
polaritons in a planar microcavity could propagate over a macroscopic scale of a fraction of 
millimeter order.
It was supposed  in Ref. \cite{Nelsen:12} that  
a BEC coherent flow formed at high enough densities of the long-lifetime polaritons 
have promoted their long-distance propagation. 
In this Article, we demonstrate that the polariton dynamics observed in Ref. \cite{Nelsen:12} is consistent
with that predicted from the  Gross-Pitaevskii equation for a polariton BEC.
Specifically, we show that under the action of an external force, a long-lifetime polariton BEC 
can propagate over the distance $\sim 0.5$ mm from the excitation spot. 
It is demonstrated  that the polariton-exciton interactions can also significantly change the spatial 
distribution of the polaritons in the BEC.
Based on the obtained results, we  propose an observation of long-lifetime polariton BEC  
formed in a graphene layer embedded into a high-Q microcavity. It is shown 
that for graphene there is an additional controlling parameter -- the energy gap in the electron and 
hole excitation spectra -- that allows one to govern the polariton BEC propagation.
The possibility to dynamically adjust the properties of the polariton BEC 
by changing the gap, for example through application of the electric field, 
makes a  microcavity with embedded graphene a promising candidate for the use 
in adaptive optical circuits.

Our article is organized in the following way. In Sec.\ II the dynamics of a Bose-Einstein condensate of long-lifetime exciton polaritons in a microcavity with an embedded semiconductor quantum well or a grapheme layer is considered based on the Gross-Pitaevskii equation. The details of the numerical simulations are presented in Sec.\ III.  Sec.\ IV presents the results of the simulations of the polarition BEC propagation in a microcavity with an embedded quantum well and an embedded graphene layer. Here, the effect of the polariton lifetime on the BEC dynamics is discussed and the comparison of the BEC propagation in a microcavity with an embedded quantum well and graphene is presented.  Finally, the conclusions follow in Sec\ V.

\vspace{-0.5cm}
\section{Dynamics of an exciton-polariton BEC in a microcavity}\label{sec:theory}
\vspace{-0.3cm}

At temperature much lower than the BEC transition temperature the dynamics of the exciton polariton BEC 
is described by the Gross-Pitaevskii equation
% \begin{equation}
%      i \hbar \frac{\partial \Psi(\bm{r},t) }{\partial t} =   - \frac{\hbar^{2}}{2 m_{\rm pol}}  
%         \Delta \Psi(\bm{r},t)  +  U(\bm{r})\Psi(\bm{r},t) +   
%     g \Psi(\bm{r},t) |\Psi (\bm{r},t)|^{2} - i\hbar \gamma \Psi (\bm{r},t) + i \hbar G(\bm{r}). \label{eq:gpnl}
% \end{equation}
\begin{eqnarray}
    & & i \hbar \frac{\partial \Psi(\bm{r},t) }{\partial t} =   - \frac{\hbar^{2}}{2 m}  
        \Delta \Psi(\bm{r},t)  +  U(\bm{r})\Psi(\bm{r},t) \nonumber \\
     & & \,\,\,\,\,\, +   g \Psi(\bm{r},t) |\Psi (\bm{r},t)|^{2} - i\hbar \gamma \Psi (\bm{r},t) + i \hbar G(\bm{r}). \label{eq:gpnl}
\end{eqnarray}
In Eq.\ (\ref{eq:gpnl}) $\Psi(\bm{r},t)$ is the condensate wave function which depends on the
coordinates  $\bm{r}=(x,y)$ in the plane of the microcavity and time $t$, $m$ is the polariton mass,
$g$ is  the effective interaction strength proportional to the polariton-polariton scattering amplitude,
$\gamma = 1/2 \tau$ is the decay rate,  $\tau$ is the polariton 
lifetime, and $ G(\bm{r})$ is the source term.
The term $U(\bm{r}) =  U_{\rm ex - pol}(\bm{r}) + U_{\rm ext}(\bm{r})$ 
describes the repulsive interaction $ U_{\rm ex - pol}(\bm{r})$
of the polaritons with those excitons in a quantum well, 
which are not coupled with the microcavity photons,
and  the  potential energy due to the external force, $U_{\rm ext}(\bm{r})$.
Importance of the interactions of polaritons with the exciton cloud
has been recently discussed in Ref. \cite{Christmann:12}.
The corresponding interaction energy  
is  
\begin{equation}
 U_{\rm ex - pol}(\bm{r}) = X_0^2 g_{\rm ex-ex} n_{\rm ex}(\bm{r}), \label{eq:uexpol}
\end{equation}
where $X_{0} \equiv X_{\bm{k}=0}$ is the Hopfield coefficient  for the exciton component in the polariton wave function 
taken at the wave vector $\bm{k}=0$,
$g_{\rm ex-ex}$ is the exciton-exciton repulsive interaction constant and 
$n_{\rm ex}(\bm{r})$ is the  exciton density in the cloud.  
The Hopfield coefficient for arbitrary wave-vector $k$ is
\begin{equation}
X_k =  \frac{1}{\sqrt{1 + \left({ \hbar \Omega_R \over  E_{\rm LP}(k)  - E_{\rm C}(k)}\right)^2}}, 
\end{equation}
where $\Omega_R$ is the Rabi splitting, 
\begin{eqnarray}
E_{\rm LP}(k) & = & {  E_{\rm C}(k) + E_{\rm X}(k) \over 2}   \nonumber\\
  & & - {1 \over 2} \sqrt{ ( E_{\rm C}(k) - E_{\rm X}(k))^2 + 4 (\hbar \Omega_R)^2 } 
\end{eqnarray}
is the energy of the lower polariton branch, $E_{\rm X}(k)$ is the exciton energy 
and $E_{\rm C}(k)$ is the cavity photon energy as functions of $k$ \cite{Ciuti:03}. 
Because the polaritons in the BEC are condensed into the state with small wave-vectors $\bm{k}$, 
the Hopfield coefficient in Eq.\ (\ref{eq:uexpol}) is taken at $\bm{k}=0$.
We approximate the exciton density in the cloud via a Gaussian profile as
$n_{\rm ex}(\bm{r})  = n_0 \exp (-|\bm{r} - \bm{r_0}|^2 / a^2)$ 
where $n_0$ is the  density at the center of the cloud and 
$a$  is the characteristic size of the cloud.
The Gaussian profile gives the required bell-like dependence for the 
exciton cloud density in the excitation area $|\bm{r} - \bm{r_0}| \sim a$ 
and also captures the sharp drop of the exciton density at large distances
$|\bm{r} - \bm{r_0}| \gg a$ \cite{Dalfovo:99}.
Thus, the polariton-exciton interaction (\ref{eq:uexpol}) is represented as
\begin{equation}
U_{\rm ex - pol}(\bm{r}) =  U_0 \exp (-|\bm{r} - \bm{r_0}|^2 / a^2), \label{eq:uexpol1}
\end{equation}
where $U_0 = X_0^2 g_{\rm ex-ex} n_0$. 
To study the effects of the polariton-exciton cloud interaction, 
we vary the interaction strength  $U_0$ 
as described below.

It has been demonstrated in recent experiments \cite{Sermage:01,Nelsen:12} 
that the microcavity polaritons can be  accelerated by an external force 
due to the spatial gradient of thickness of the microcavity.
In this approach, the average force acting upon a polariton wave packet at the point $\bm{r}$ 
is  $\bm{F}(\bm{r}) = - \nabla E_{\rm C}(\bm{r})$, where $E_{\rm C}(\bm{r})$ is energy of the polariton band 
taken at the in-plane  vector of the polariton $\bm{k}=0$ \cite{Sermage:01}.
In particular, in experiments \cite{Sermage:01,Nelsen:12} the microcavity width, and hence the energy $E_C(\bm{r})$,
has been a linear function a spatial coordinate (the wedge-shape geometry of the  microcavity) that resulted in a 
constant force $\bm{F}(\bm{r}) = \bm{F}_0$ acting upon the polaritons.
To describe the effect of the force,  we set the corresponding potential energy as  
\begin{equation}
U_{\rm ext}(\bm{r}) = - F_0 x. \label{eq:Uf}
\end{equation}

To explore the polariton BEC dynamics in a macroscopically 
large sample we consider the case where the source of polaritons,  
$G(\bm{r})$, is localized in space \cite{Roumpos:12}. 
Within the used approximation, the source term is represented as 
$G(\bm{r}) = G_0 \exp (-|\bm{r} - \bm{r_0}|^2 / a^2)$ 
with the same characteristic size $a$
as for the exciton cloud size.

Following the conditions of the experiments \cite{Sermage:01,Nelsen:12}, 
we varied $U_0$ from 0 to  10 meV and $F_0$ from 0 to 13~meV/mm. For the 
long-lifetime polaritons  in a microcavity with the quality factor $Q \sim 10^6$ 
with an embedded semiconductor QW we set $\tau = 100$ ps \cite{Nelsen:12}, 
$m = 10^{-4} m_0$ where $m_0$ is the free  electron mass \cite{Carusotto:13}, 
and $g = 10^{-2}$ meV$\mu$m$^2$  \cite{Amo:09}.
 
 \begin{figure}[t]\vspace{-.4cm} 
\centerline{\includegraphics[width=9.cm]{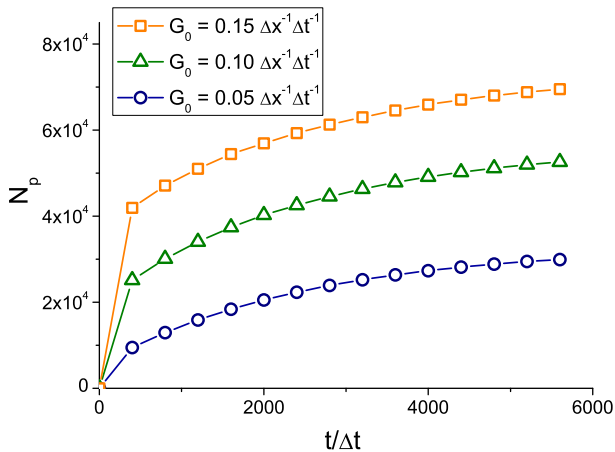}}
\vspace{-.5cm}
\caption{\label{fig:npart} Time dependence of the total 
number of polaritons in the BEC for $F_0 = 13$ meV/mm and $U_0 = 3$ meV for different 
pumping rates $G_0$. }\vspace{-0.3cm}
\end{figure}
 
Formation of the polariton BEC in gapped graphene has recently been 
predicted in Ref.\ \cite{Berman:12c}. The polariton
effective mass in graphene is 
\begin{equation}
m = 2 (m_{\rm ex}^{-1} + c L_c / \sqrt{\varepsilon} \pi \hbar) ^{-1}, \label{eq:graphm}
\end{equation}
where $ m_{\rm ex}$ is the exciton effective mass in graphene, $L_c$ is the length of the microcavity, 
$\varepsilon$ is the dielectric constant of the microcavity, and $c$ is the speed of light in vacuum. 
In the present work we  consider the case of zero detuning where the cavity 
photons and the excitons in graphene are in resonance at $\bm{k}=0$. 
In this case the length of the microcavity is chosen as \cite{Berman:12a}
\begin{equation}
 L_c = \frac{\hbar \pi c}{(2 \delta - V_0 + C / \delta^2)\sqrt{\varepsilon}}, \label{eq:lc}
\end{equation}
where $V_0 = e^2 / 4 \pi \varepsilon \varepsilon_0 r^{\prime}$, $C = (\hbar e v_F)^2 / 8 \pi \varepsilon \varepsilon_0 r^{\prime 3}$, 
$\delta$ is the gap energy in the quasiparticle spectrum, 
and $v_F$ is the Fermi-velocity of the electrons in graphene.  
The parameter $r^{\prime}$ is found from the following equation,
\begin{equation}
2 \delta^2(2 \delta - \hbar \omega) r^{\prime 3} - 2 D \delta^2 r^{\prime 2} + D(\hbar v_F)^2 = 0,
\end{equation}
where $D = e^2 / 4 \pi \varepsilon \varepsilon_0$. For dipolar excitons in GaAs/AlGaAs coupled quantum wells, 
the energy of the recombination peak is $\hbar \omega = 1.61$ eV \cite{Negoita:99}.
We expect similar photon energies in graphene. However, its exact value depends on the graphene dielectric 
environment and substrate properties though. The exciton effective mass in Eq. (\ref{eq:graphm}) is given by
\begin{equation}
  m_{\rm ex} = \frac{2 \delta^4}{Cv_F^2}. \label{eq:mex}
\end{equation}
The polariton-polariton interaction strength is  \cite{Berman:12c,Ciuti:03}
\begin{equation}
g = {3 e^2 a_B \over 8 \pi \varepsilon_0 \varepsilon}, \label{eq:graphg}
\end{equation}
where $a_B = 2 \pi \varepsilon_0 \varepsilon  \hbar^2 / m_r e^2$ is the two-dimensional Bohr radius of the exciton, $m_r$
is the exciton reduced mass.

Formation of the excitons in graphene requires a gap in the electron and hole excitation 
spectra, which can be created, for example, by external electric field
\cite{Kuzmenko:09, Mak:09, Zhang:09}. We do not restrict our consideration to a specific 
mechanism of the gap formation.
As follows from Eqs.~(\ref{eq:graphm})--(\ref{eq:graphg}),  
both the polariton mass $m$ and the interaction strength $g$ in gapped graphene depend  on the gap energy $\delta$.
In our studies, we varied  $\delta$ from 0.1 eV to 0.5 eV that are the representative values for graphene.
The polariton lifetime is mainly defined by the leakage of the photons 
through the mirrors and only weakly depends on the exciton 
lifetime \cite{Carusotto:13,Nelsen:12}. 
Therefore, we use the same  polariton lifetime  $\tau=100$ ps for graphene and for a semiconductor 
quantum well.

\vspace{-0.7cm}
 \begingroup
\let\clearpage\relax
 \section{Numerical simulations}\label{sec:numerics}
\endgroup

To study the polariton BEC dynamics, we numerically integrated Eq.\ (\ref{eq:gpnl}).
At the initial moment  $t=0$ we set $\Psi(\bm{r},0)=0$.
As is demonstrated in Sec.\ \ref{sec:results} below, the long-lifetime polaritons  
accelerated by the external force can propagate over distances up to
$l_{\rm max} \sim 0.5$ mm from the excitation spot. The characteristic wavelength of the polariton wave packet at the 
distance $l \approx l_{\rm max}$  can be estimated quasiclassically  as 
$\lambda_{\rm min} \approx 2 \pi \hbar / \sqrt{2 m (F_0 l_{\rm max} + U_0)}$ that gives
$\lambda_{\rm min} \approx 9.5 \times 10^{-7}$m for   $F_0 = 13$~meV/mm and $U_0 = 10$~meV.
At shorter distances  $l < l_{\rm max}$ one has $\lambda > \lambda_{\rm min}$.  
To  resolve the oscillations of the polariton wavefunction  
with the wavelengths $\lambda \geq \lambda_{\rm min}$, 
we choose the numerical grid spacing 
$\Delta x = 1.5 \times 10^{-7}$ m. The dimensions of the numerical box are
$L_x \times L_y = 6000 \times 3000 \, (\Delta x)^2$.
The long side of the numerical box is arranged parallel to the direction of the constant force $\bm{F}_0$.
We use the periodical boundary conditions at the numerical box boundaries.
The unit of time is equal to $\Delta t = 2 m (\Delta x)^2 / \hbar = 
3.9 \times 10^{-14}$~s.
We numerically integrated Eq.\ (\ref{eq:gpnl}) over time
with the 4$^{\rm th}$ order Runge-Kutta scheme at the time step $h = 10^{-2} \Delta t$. 
An explicit 4$^{\rm th}$ order accuracy method has been used to represent the Laplacian operator in Eq. (\ref{eq:gpnl}).
We took advantage of the graphical processing unit CUDA programming \cite{Nvidia} that allowed us 
to achieve $\sim 20 \times$ increase of the performance  with respect to the shared-memory parallel 
version of the code. 
This allowed us to perform the simulations for this macroscopic system
within practically realizable time intervals.
The overall computation time for a single run was $\sim 10$ hours on  NVIDIA Tesla S2050  card.

%\vspace{0.5cm}
\section{Propagation of a BEC of long-lifetime polartions in a microcavity}\label{sec:results}
\vspace{-0.2cm}
\subsection{Polartion BEC in a microcavity with an embedded quantum well}\label{sec:qw}
\vspace{-0.3cm}

In this Section we focus on the propagation of a BEC of long-lifetime polaritons
in a microcavity with an embedded semiconductor QW.
Fig.~\ref{fig:npart} shows the time dependence of the total number of polaritons 
$N_p(t) = \int d^2 \bm{r} |\Psi(\bm{r},t)|^2$
after the source it turned on. 
It follows from Fig.~\ref{fig:npart} that after the system approaches a steady state 
at $t \sim 6 \times 10^3 \Delta t \approx 230$ ps, for the pumping rate
$G_0 =  0.1 \Delta x ^{-1} \Delta t ^{-1}$ the total number of polaritons 
is $N_p \sim 5.3 \times 10^4$ that is a representative value 
for experiments with exciton polaritons in a semiconductor microcavity \cite{Nelsen:12}.
Below we used this  pumping rate in our simulations.

\begin{figure}[b] \vspace{-0.2cm}  
\includegraphics[height=3.5cm]{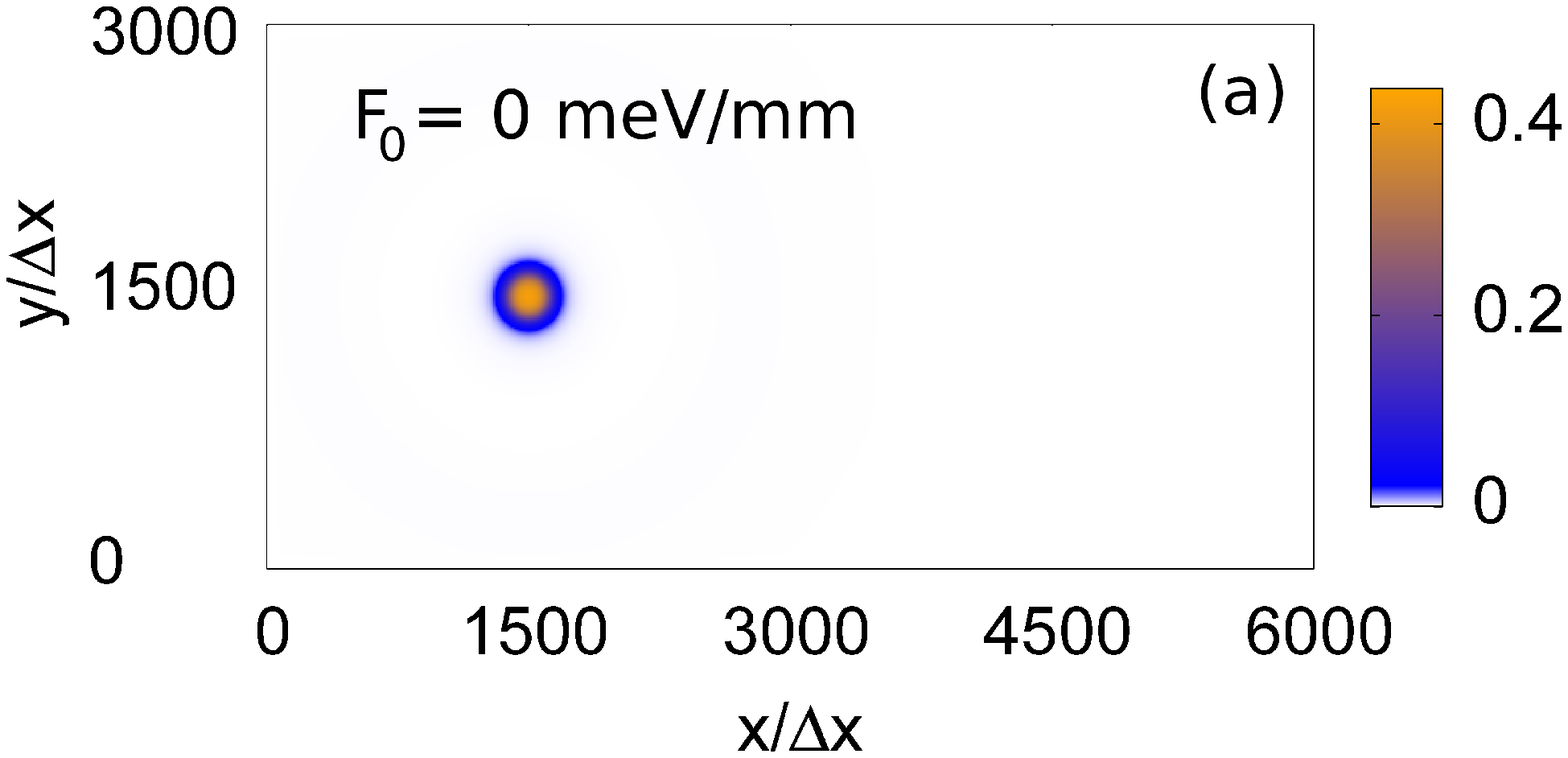}
\includegraphics[height=3.5cm]{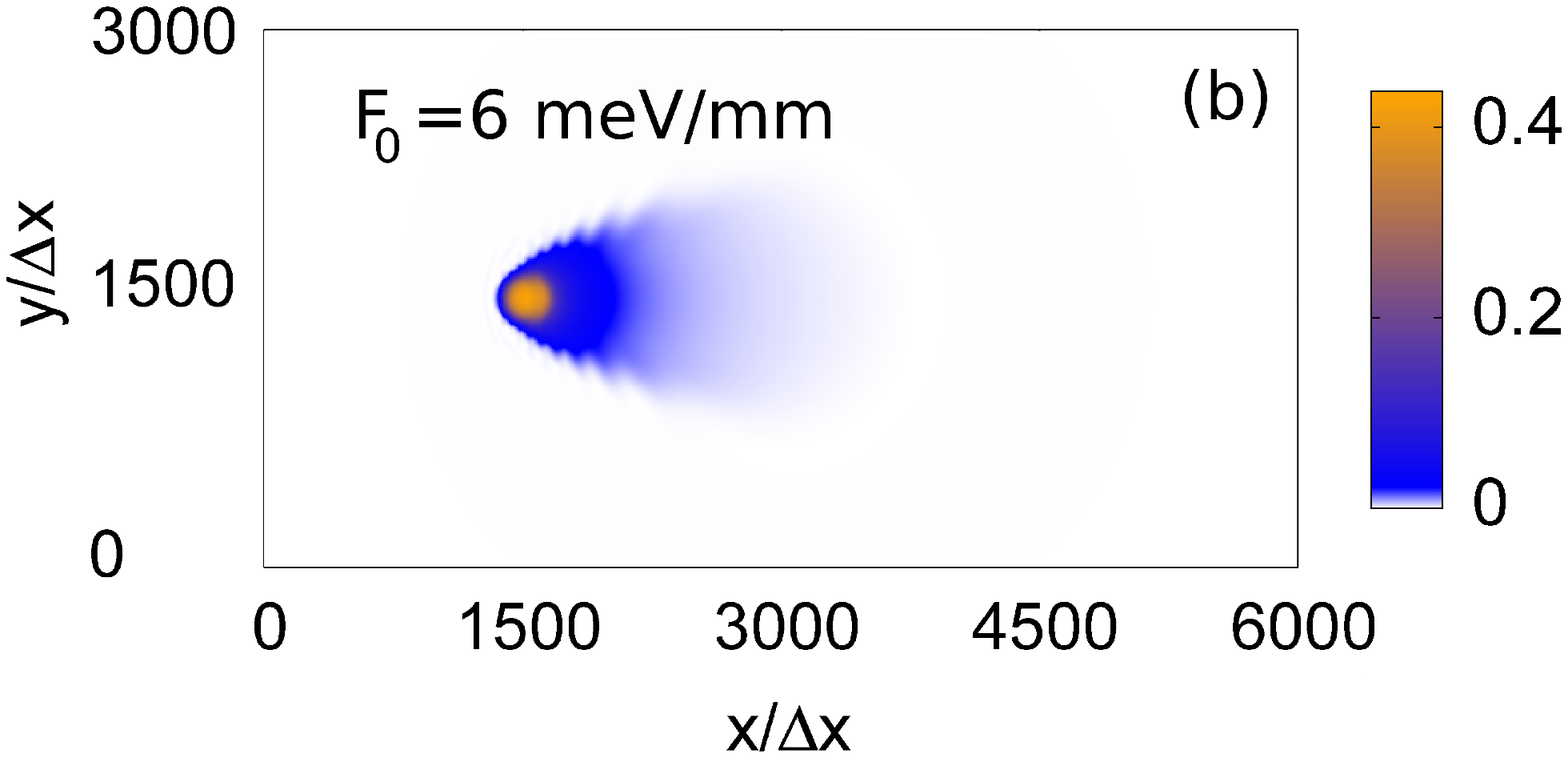}\\
\includegraphics[height=3.5cm]{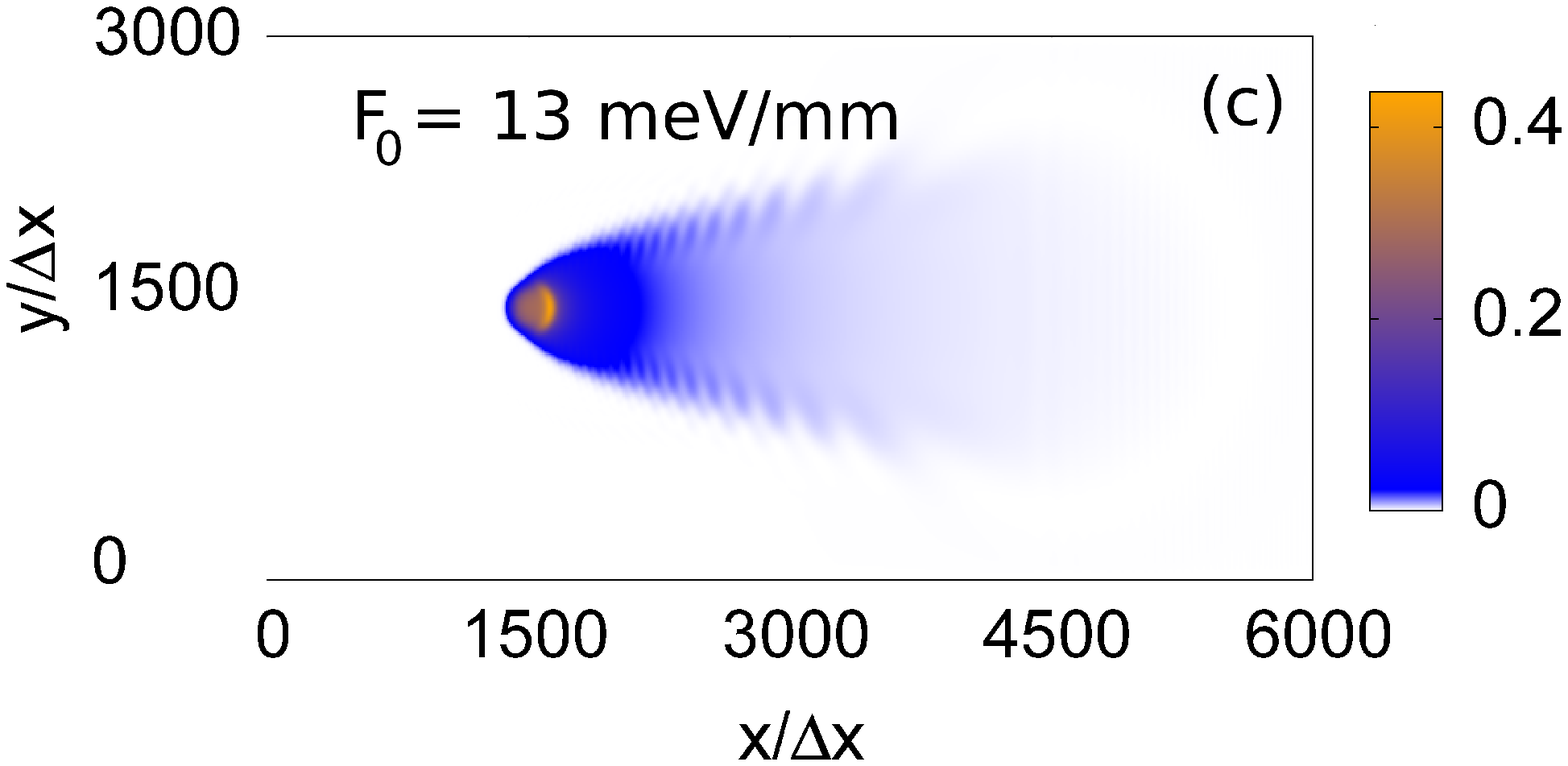}\\
\hspace{-.5cm}%
\includegraphics[width=8cm]{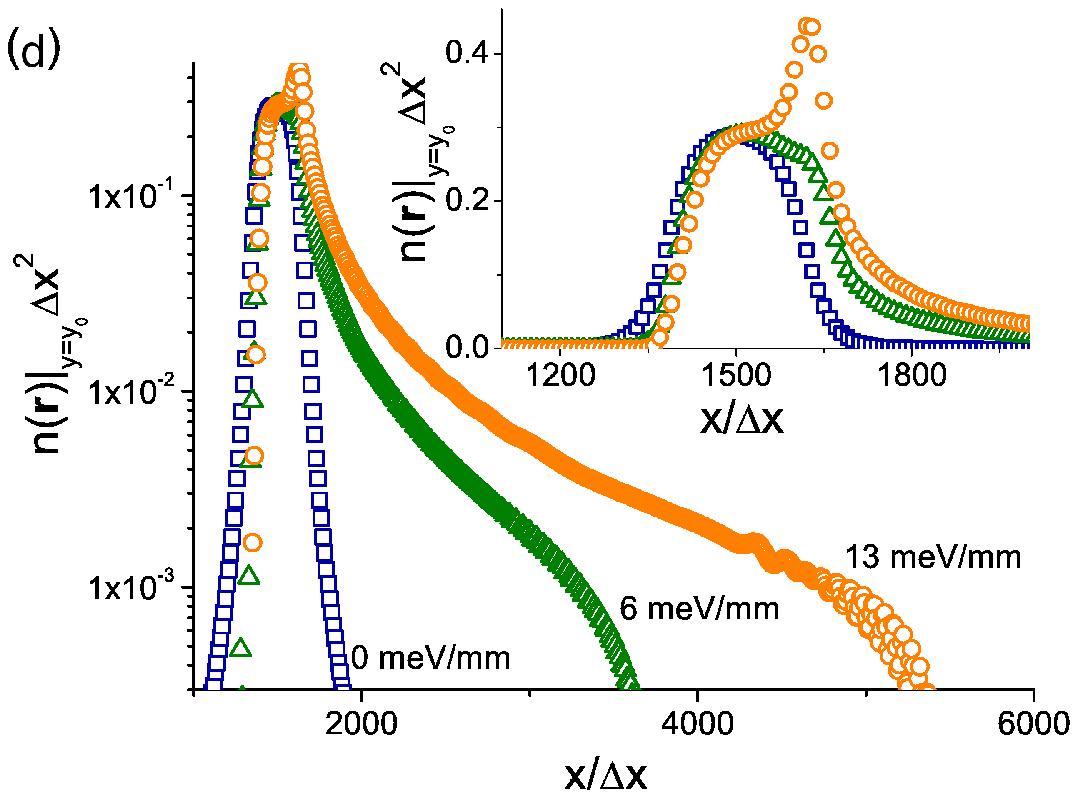}
\vspace{-0.5cm}
\caption{\label{fig:dens}   (a-c) Evolution of 
a trace of  the long-lifetime  polariton BEC with increasing the force $F_0$ for 
$t = 4800 \Delta t$ and $U_0 = 3$ meV.
The colorbars show the dimensionless BEC density $n(\bm{r})\Delta x^2$. 
(d) Evolution of a steady-state polariton density at $y=y_0$ for $U_0 = 3$ meV
with increasing $F_0$  shown in a logarithmic scale and in a linear scale 
in the region around the excitation spot (inset).}%\vspace{-0.7cm}
\end{figure}

Propagation of long-lifetime polariton BEC in the quantum well plane is shown in Fig.~\ref{fig:dens}. 
The center of the excitation spot is positioned at 
$\bm{r}_0 = (x_0, y_0)$ where $x_0 = y_0 = 1500 \Delta x$. 
The evolution of  the long-lifetime polariton
BEC distribution with increasing of the force $F_0$ is given in Fig.\ \ref{fig:dens}a-c.
Fig.~\ref{fig:dens}a shows
that in the absence of the force the polariton spot is symmetric 
in the $(x,y)$ plane and has a characteristic size of $\approx 500 \Delta x = 75$~$\mu$m 
that is, $\sim 7.5\, a$. With the increase of the force a polariton  
``trace'' is formed  and at $F_0=13$ meV/mm 
the length of the trace reaches $\sim 3.5 \times 10^3 \Delta x \approx 500$~$\mu$m (Fig.\ \ref{fig:dens}c).
To characterize the steady-state polariton distribution in details
we plot  in Fig.\ \ref{fig:dens}d  the polariton BEC density at $y=y_0$ that is, a cross-section of 
Fig.\ \ref{fig:dens}a-c along the line passing through the center of the excitation spot.
The inset in Fig.\ \ref{fig:dens}d shows that for the force $F_0=13$ meV/mm, 
the maximum of the polariton density is shifted from the center of the excitation spot
$x_0$ to $x \approx 1620 \Delta x$. 

\begin{figure}[t] 
\includegraphics[width=4.cm,height=3.5cm]{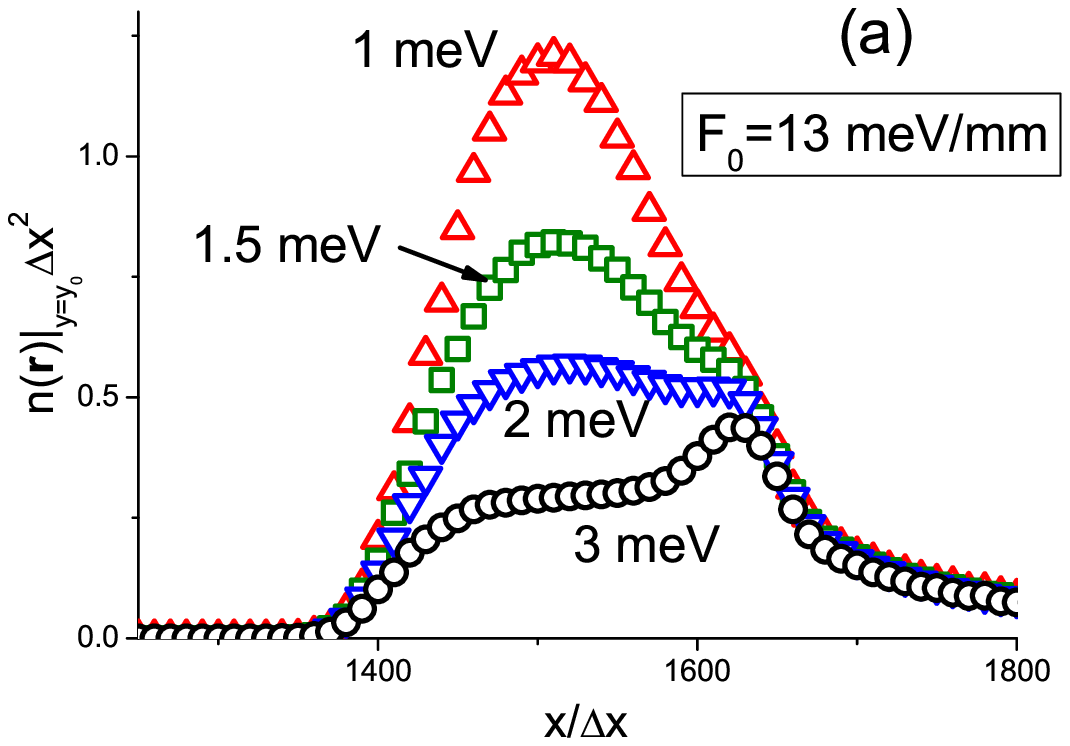}\includegraphics[height=3.5cm]{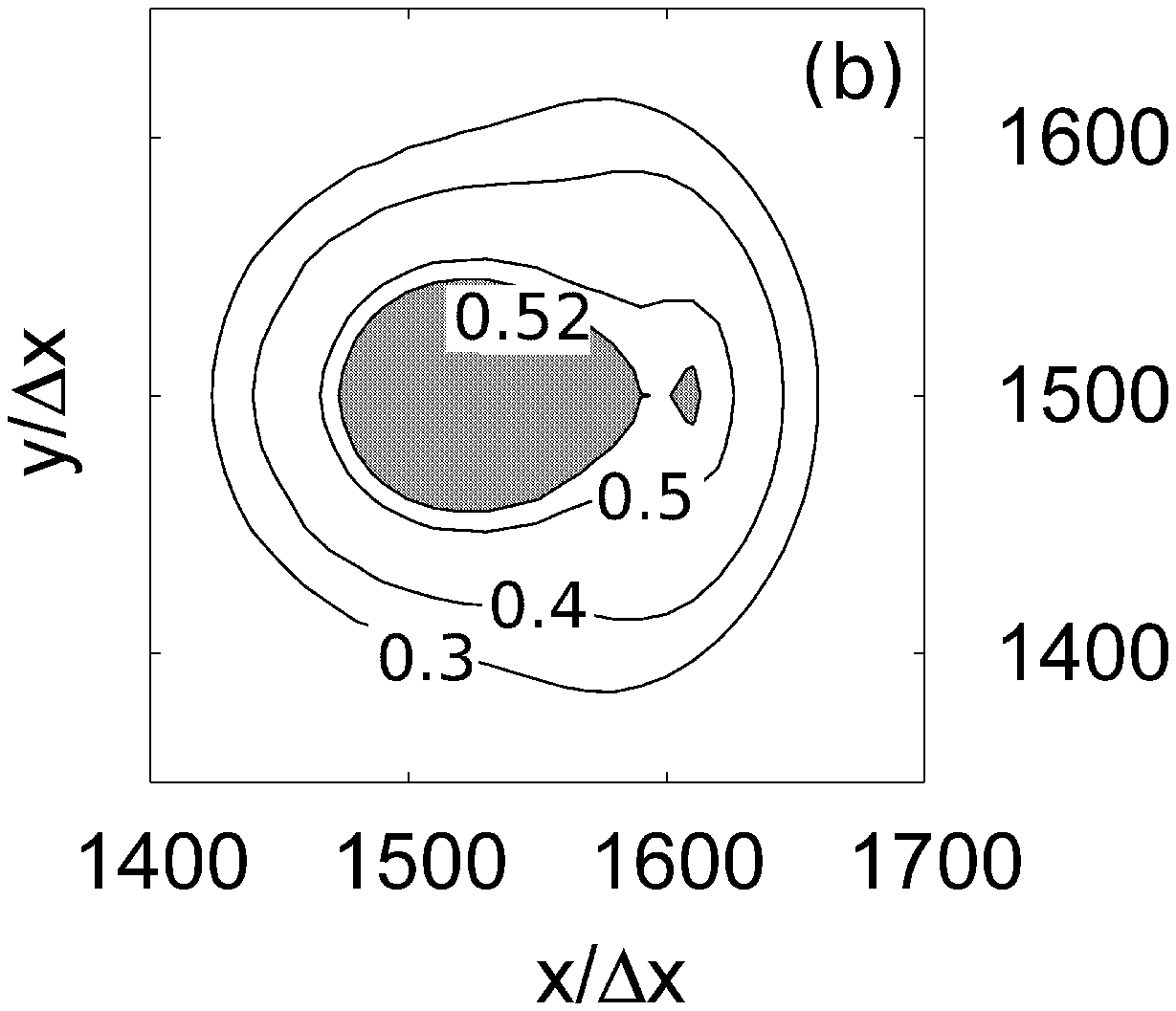}
\includegraphics[height=3.5cm]{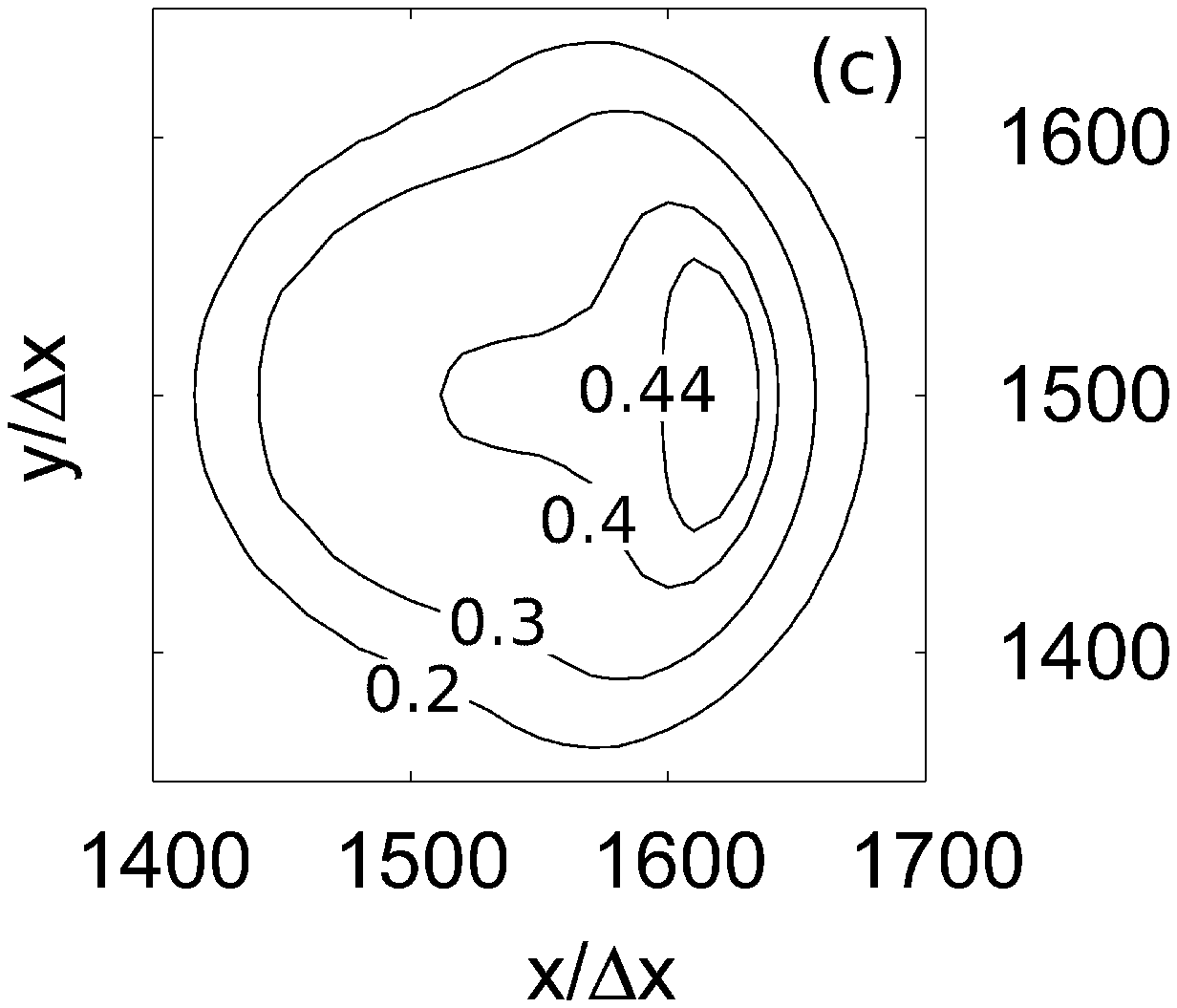}\includegraphics[height=3.5cm]{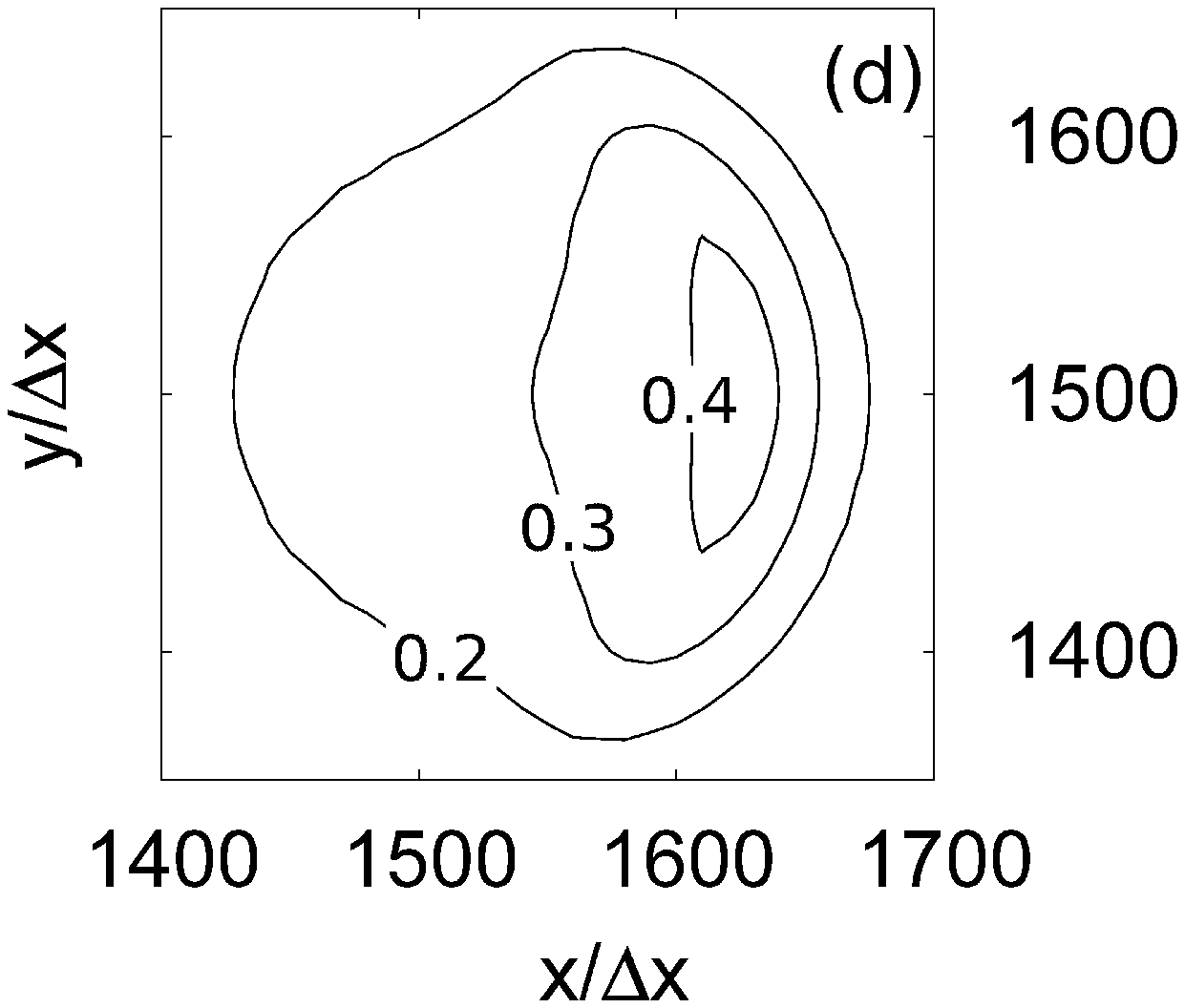}
\vspace{-0.3cm}
\caption{\label{fig:upol}  (a) Changes in the polariton density at $y=y_0$   
with the increase of $U_0$ from 1 meV to 3 meV
%maximum effective polariton-uncoupled excitons interaction strength: 
%$U_0$ =  1 meV ($\bigtriangleup$),  1.5 meV ({\scriptsize $\bigcirc$}),  2 meV ($\Box$), and  3 meV ($\bigtriangledown$).
for $t=5.8\times 10^3 \Delta t$.
(b-d): Contour plots for the polariton density  for $U_0 =$  2 (b),  2.5 (c), and  
3 meV (d); other parameters are the same as in plot (a). The contour lines are labeled 
with the values of the density $n(\bm{r}) \Delta x^2$. 
In plot (b) the area with $n(\bm{r}) \Delta x^2 \geq 0.52$ is gray-shaded.}%\vspace{-0.7cm}
\end{figure}

To prove that the shift of the density maximum  is caused by the 
interaction of the polaritons with the exciton cloud, 
we investigated evolution of the polariton 
density with changes of the maximum effective interaction strength $U_0$.
Fig.~\ref{fig:upol}a demonstrates changes in the dimensionless polariton density along the line $y=y_0$ with the rise 
of $U_0$ from 1 meV through 3 meV. It is seen that 
the position of the maximum of the density is shifted
from $x \approx 1500 \Delta x$ at $U_0 = 1$ meV to $x \approx 1620 \Delta x$ at $U_0 = 3$ meV.
Fig. \ref{fig:upol}b-d shows the contour lines for 
the  polariton density $n(\bm{r})$  for different $U_0$.
From Fig.\ \ref{fig:upol}b it is evident that at $U_0 = 2$ meV and $y=y_0=1500 \Delta x$
the spatial polariton distribution 
has {\it two maxima} positioned at $x\approx 1500 \Delta x$ and $1620 \Delta x$. 
While increasing $U_0$ to 2.5 meV the left maximum disappears and the right maximum 
becomes more pronounced, as seen in Fig.\ \ref{fig:upol}c. With the further increase of $U_0$, 
most polaritons are concentrated around  
the maximum at $x\approx 1620 \Delta x$ (Fig.\ \ref{fig:upol}d).

\begin{figure}[h] 
\includegraphics[height=4.4cm]{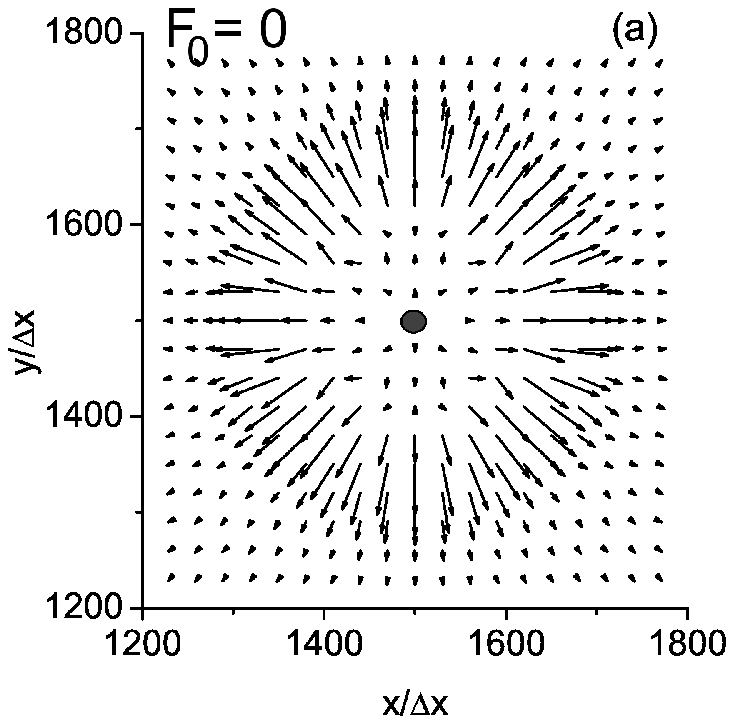}\hspace{-0.5cm}
\includegraphics[height=4.4cm]{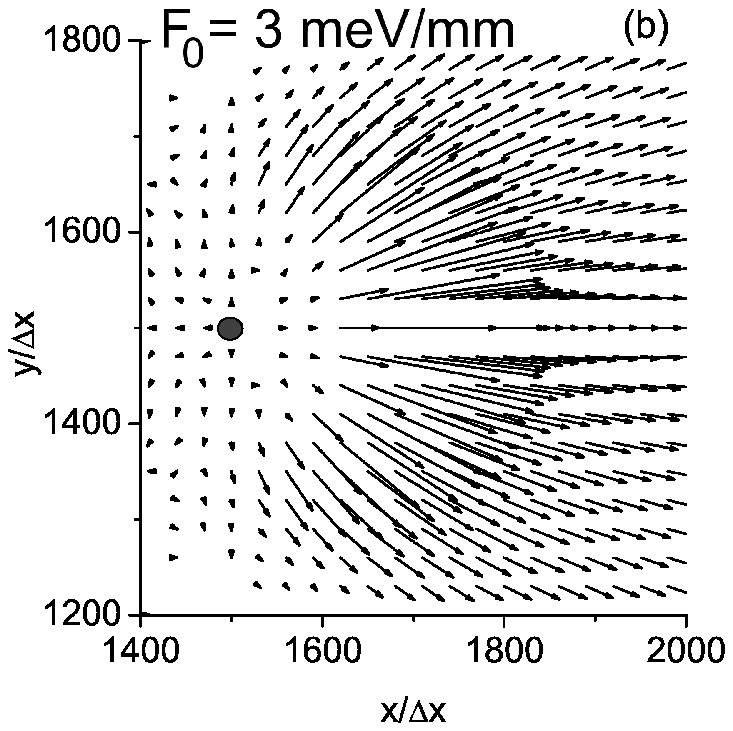}\vspace{-0.5cm}
\caption{\label{fig:j} Polariton flux $\bm{j}$ for 
$U_0 = 3$ meV, $t=5.6 \times 10^3 \Delta t$. 
The length of the arrows is proportional to the magnitude of the flux  $|\bm{j}|$. The  point 
at $\bm{r}_0 = (1500 \Delta x, 1500 \Delta x)$ marks the center of the pumping spot. }%\vspace{-0.7cm}
\end{figure}

To clarify the reasons for the shift of the density maximum while $U_0$ is increased,
we compare in Fig.\ \ref{fig:j} the distribution of the polariton flux in a steady state
$\bm{j}(\bm{r}) = ({\hbar / 2 i m}) \left( \Psi^*(\bm{r}) \nabla \Psi(\bm{r}) - \Psi(\bm{r}) \nabla \Psi^*(\bm{r})\right)$
for different forces.
For $F_0=0$ the  flux $\bm{j}(\bm{r})$ is distributed symmetrically around
the center of the excitation spot. 
 As is seen in Fig.\ \ref{fig:j}a  the magnitude of the flux, $j$, reaches its maximum 
at the distance $\sim 120 \Delta x$ from the excitation spot center $\bm{r}_0$. 
 The presence of the maximum of  $j$ at certain distance from the point $\bm{r}_0$ 
can be understood if one considers the creation and propagation of polariton wave packets
in the potential energy profile $U_{\rm ex - pol}(\bm{r})$. 
The polaritons are accelerated due to the force $\bm{F}_{\rm ex - pol}(\bm{r}) = - \nabla U_{\rm ex - pol}(\bm{r})$ 
and hence, their velocity $v$ gradually rises with the distance from the center. 
At large distances, the velocity $v$ tends to a constant because the force $\bm{F}_{\rm ex - pol}(\bm{r})$ vanishes.
On the other hand, the polariton density $n(\bm{r})$ 
falls at large distances from $\bm{r}_0$ due to spreading of the polaritons over the sample. 
In effect, the flux $j = n v$ reaches the maximum at certain 
distance $\sim a$ from the excitation spot center. 

In the presence of a force $\bm{F}_0$, the polariton flux distribution  acquires
a strongly asymmetric form.   The result of the simulations for $F_0 = 3$ meV/mm and 
$U_0 = 3$ meV is shown in Fig.\ \ref{fig:j}b. 
 The simulations for other forces $F_0$ and energies $U_0$ resulted in the polariton flux distributions, 
which were similar to that shown in  Fig.\ \ref{fig:j}b.
In this case, most of the polaritons move  in the direction of the force $\bm{F}_0$. 
This results in the shift of the maximum of the polariton density 
from the spot center, in agreement with Fig.\ \ref{fig:npart}d and  Fig.\ \ref{fig:upol}.
It is worth noting that at distances $\sim a$ from the excitation spot center 
there is a finite flux directed opposite to the external force $\bm{F}_0$. 
However, these polaritons eventually  flow around the excitation spot and, at large distances from the spot,
move in the direction of the force in agreement with the observations in Ref. \cite{Nelsen:12}.

% \begin{figure}[t] %\vspace{-0.5cm}
% \includegraphics[width=8.5cm]{grapheneonly.eps}%\vspace{-0.5cm}
% \caption{\label{fig:graphene}  Variation of the density of a 
% long-lifetime polariton BEC  in graphene  for $F_0=13$ meV/mm and $U_0 = 5$ meV
% with rising the gap energy $\delta$.
% %and the density of a short-lifetime polariton BEC ($\tau=3$~ps) 
% %for a GaAs-based microcavity for $F_0=13$ meV/mm and 
% %$U_0=5$ meV (open circles).
% } %\vspace{-0.5cm}
% \end{figure}

\begin{figure}[t] %\vspace{-0.5cm}
\includegraphics[width=8.5cm]{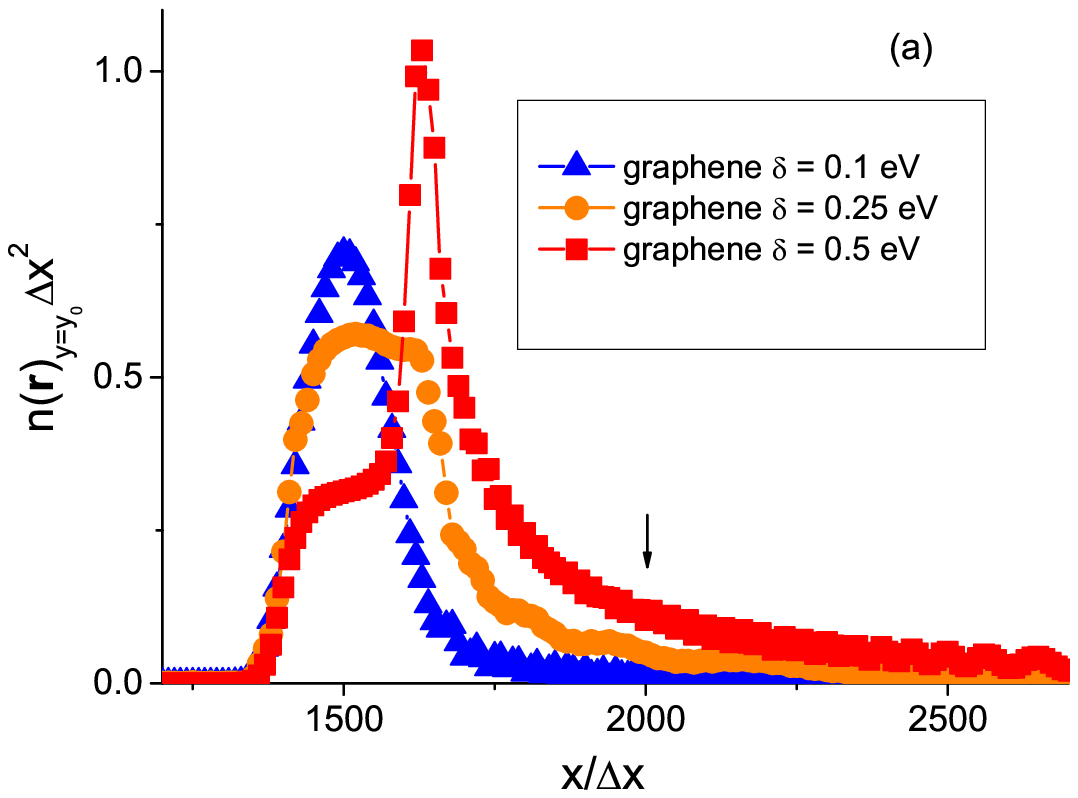}\vspace{-0.3cm}
\includegraphics[width=8.5cm]{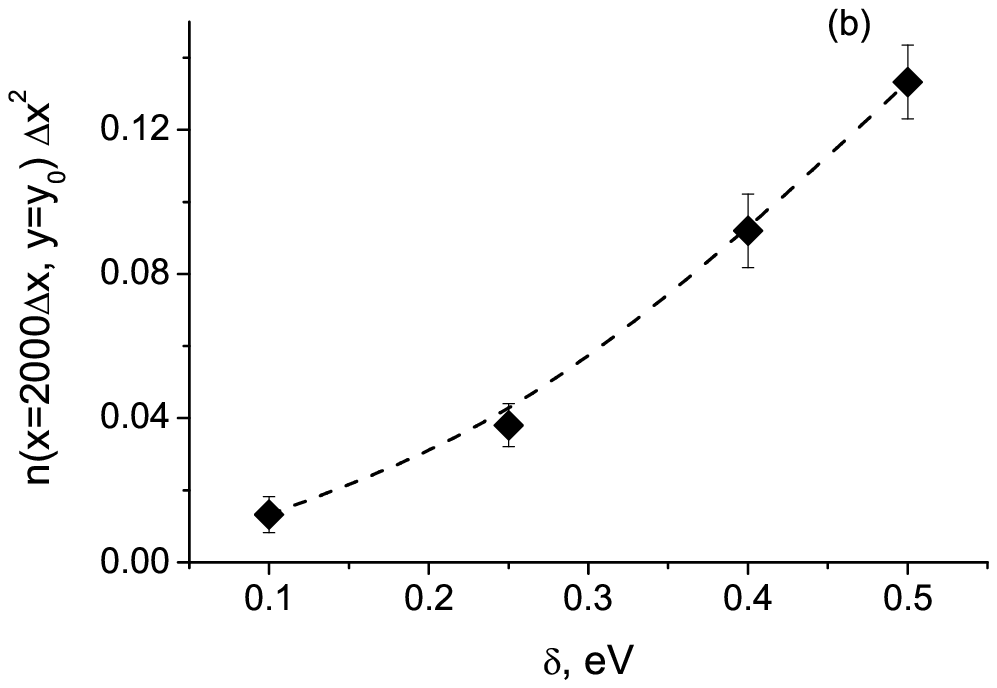}\vspace{-0.5cm}
\caption{\label{fig:graphene}  (a) Variation of the density of a 
long-lifetime polariton BEC  in a microcavity with embedded graphene 
with rising the gap energy $\delta$. (b) Points: dependence of the polariton BEC density at the point
$\bm{r} = (2000\Delta x, 1500 \Delta x)$ (marked by a vertical arrow in plot (a)) with changing $\delta$. 
 Vertical errorbars show uncertainty due to small 
time-dependent oscillations of the polariton density about the mean values.
The dashed curve is a guide to the eye.
$F_0=13$ meV/mm and $U_0 = 5$ meV for both plots (a) and (b).
} \vspace{-0.3cm}
\end{figure}

\vspace{-0.5cm}
\subsection{Polariton BEC dynamics in a microcavity with embedded graphene}\label{sec:graphene}
\vspace{-0.3cm}
Within the same approach, we also studied the propagation of polaritons in 
a semiconductor  microcavity with gapped graphene embedded into it.  
First, we consider the case of the GaAs-based microcavity, for which we set $\varepsilon = 13$.
The effective mass $m$ of the polaritons and the polariton-polariton interaction strength $g$
have been found from Eqs. (\ref{eq:graphm})--(\ref{eq:graphg}) for different values of the gap energy $\delta$.
The results of the simulations for graphene in Fig.~\ref{fig:graphene}a shows
that the polariton density in the trace  gradually increases with rising the gap energy. 
 To characterize the increase of the polariton BEC density in the trace with changing the gap energy 
we present  in Fig.\ \ref{fig:graphene}b the dependence of the dimensionless polariton density 
at the point $\bm{r}_1 = (2000\Delta x, 1500 \Delta x)$  on the gap energy $\delta$. 
The point $\bm{r}_1$ is positioned at the distance of  $500 \Delta x \approx 75$ $\mu$m  from the excitation spot center $\bm{r}_0$.
The position of the point $\bm{r}_1$ is marked by a vertical arrow in Fig.\ \ref{fig:graphene}a.
 Fig.~\ref{fig:graphene}b shows that the polariton density at the point $\bm{r}_1$ increases in $10.2 \times$, from  
$n \approx 0.013 \Delta x^{-2}$ to $n \approx 0.133 \Delta x^{-2}$,
with changing $\delta$ from 0.1 eV to 0.5~eV. 
Additionally, the shift of the  density maximum
in the direction of the external force $\bm{F}_0$ is more pronounced 
for larger $\delta$, as shown in Fig.~\ref{fig:graphene}a.

\begin{figure}[t] %\vspace{-0.5cm}
\includegraphics[width=8.5cm]{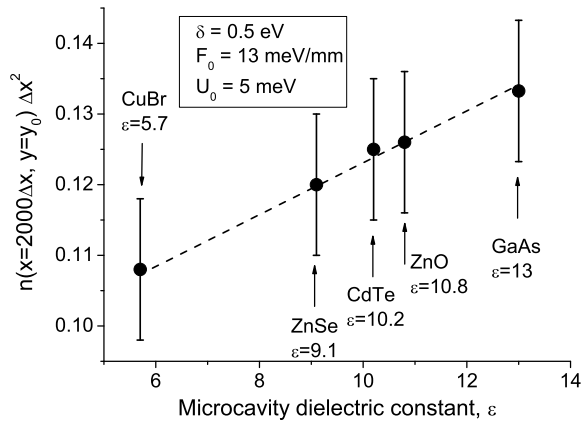}\vspace{-0.5cm}
\caption{\label{fig:eps}  Dependence of the  polariton
BEC density  in a microcavity with embedded graphene 
at the point $\bm{r}_1 = (2000\Delta x, 1500 \Delta x)$ 
on the microcavity dielectric constant  $\varepsilon$.
The dashed line is a guide to the eye.} \vspace{-0.5cm}
\end{figure}

The microcavity can be synthesized from different semiconductor materials. 
To study the effect of the microcavity material 
we investigated the propagation of the long-lifetime polariton 
BEC  in graphene in microcavities formed by semiconductor materials  CuBr, ZnSe, CdTe, and ZnO \cite{Berger:97} that is,
by those materials, which were utilized for the experimental studies of microcavity polaritons 
\cite{Kelkar:97,Dang:98,Saba:01,Zamfirescu:02,Kasprzak:06,Christopoulos:07,Manni:11,Kawase:12,Nakayama:12}.
 According to Eqs.\ (\ref{eq:graphm})--(\ref{eq:graphg}),
changes in the dielectric constant $\varepsilon$  of the microcavity result in the variations of the 
microcavity length $L_c$, for which the cavity photons and excitons in graphene are in the resonance, as well as in
the changes in the polariton effective mass $m$ and the polariton interaction strength $g$.
It was found that the BEC density distributions for CuBr ($\varepsilon = 5.7$),
ZnSe ($\varepsilon = 9.1$), CdTe ($\varepsilon = 10.2$) and ZnO ($\varepsilon = 10.8$)  microcavities
are qualitatively similar to those obtained for the microcavity synthesized from GaAs and
considered above in this Section. To characterize the details of the changes in the polariton BEC distribution in those microcavities, 
we plot in Fig.\ \ref{fig:eps}  the dependence of the dimensionless density of the condensate,
$n(\bm{r}) \Delta x^2$, at the point $\bm{r}_1 = (2000\Delta x, 1500 \Delta x)$, 
similarly to that shown in Fig.\ \ref{fig:graphene} for a GaAs-based microcavity.
It is seen in  Fig.\ \ref{fig:eps} that the polariton density at the distance $\sim 75$ $\mu$m from the excitation spot center
gradually decreases with the decrease of the dielectric constant of the microcavity material.
However, this variation is  slow: the change of the dielectric constant from $\varepsilon = 13$ (GaAs)
to 5.7 (CuBr) only results in $\sim 19$\% decrease of the BEC density from $n \approx 0.133  \Delta x^{-2}$
to  $n \approx 0.108  \Delta x^{-2}$.  Thus, the BEC density in a polariton trace in a microcavity with embedded %gapped 
graphene at large distances from the excitation spot is higher for the microcavity with high dielectric constant.

{\marko In Ref.\ \cite{Berman:12c} it was shown that the Kosterlitz-Thouless transition temperature $T_c$ for 
polaritons in a microcavity with embedded graphene increases with the decrease of the microcavity 
dielectric constant $\varepsilon$. On the other hand, from the consideration above it follows  that 
the increase of $\varepsilon$ leads to higher BEC densities at a large distance from the excitation spot. 
These conclusions are consistent with each other because
Ref.\ \cite{Berman:12c} describes the redistribution of the polaritons in the superfluid between
the normal and superfluid components at finite temperature, whereas in the present work we consider the dynamics of the system
at  a given total number of the polaritons in the BEC in the low temperature limit $T \ll T_c$.
}

\vspace{-0.5cm}
\subsection{Effect of the polariton lifetime on the BEC dynamics}\label{sec:short}
\begin{figure}[t] %\vspace{-0.5cm}
\includegraphics[width=8.5cm]{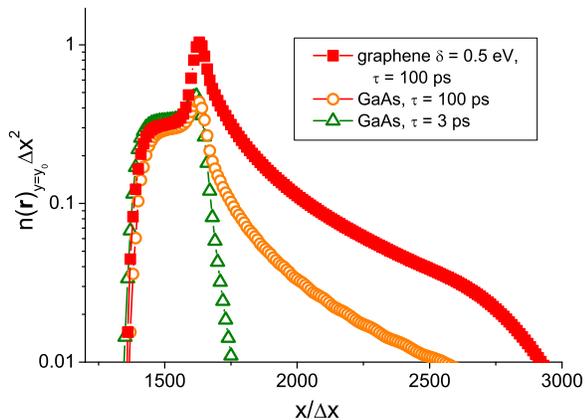}\vspace{-0.5cm}
\caption{\label{fig:short}  Density of a long-lifetime polariton BEC in graphene for 
$U_0 = 5$ meV and $\delta=0.5$ eV, in a GaAs-based microcavity for $U_0=3$ meV, and 
of short-lifetime polariton BEC  in a GaAs-based microcavity with $\tau = 3$ ps for $U_0=5$ meV. 
The force is $F_0=13$ meV/mm.
} \vspace{-0.5cm}
\end{figure}

\vspace{-0.5cm}
Finally, to demonstrate the effect of the polariton  lifetime on the BEC dynamics, we studied the BEC
spreading in a semiconductor quantum well in a microcavity
for traditional, short-lifetime polaritons, for which we took $\tau = 3$ ps \cite{Amo:09}.
{\markg In Fig.\ \ref{fig:short} we compare the BEC density for the short-lifetime 
polaritons in a microcavity with GaAs QW with that for the long-lifetime polaritons  
in a microcavity with GaAs QW and with embedded graphene at $\delta = 0.5$ eV.  
It is clearly seen that for the short-lifetime polaritons, even in the presence of an external force, 
the BEC is mostly located in the region where it is directly excited by the external pumping
and its density rapidly decreases at large distances from the excitation spot. 
Fig.\ \ref{fig:short} also shows that the long-lifetime polariton BEC density in graphene 
with $\delta = 0.5$ eV  is higher than 
that in the GaAs quantum well in a microcavity at large distances from the excitation spot.}

\vspace{-0.5cm}
{\markg 
\subsection{Comparison of the BEC propagation in microcavity with an embedded quantum well and graphene}
\vspace{-0.3cm}
From the results obtained in Sec.\ \ref{sec:qw} -- \ref{sec:short} it follows that 
the propagation dynamics of the long-lifetime polariton BEC in a microcavity with an embedded GaAs quantum well and
in a microcavity formed by} {\markg  GaAs, ZnSe, CdTe, and CuBr with an embedded graphene layer are quantitatively
similar to each other. However, 
the polariton BEC density in the trace for a GaAs-based microcavity with embedded graphene at $\delta = 0.5$ eV is higher than 
that for a microcavity with GaAs QW, as is demonstrated in  Fig.~\ref{fig:short}. 
The BEC density in the microcavity with embedded graphene sharply increases 
with the rise of the gap energy $\delta$, as shown in Fig.\ \ref{fig:graphene}b.
This allows one to utilize the gap energy in graphene as a parameter that controls the polariton  BEC propagation in a 
microcavity.  The polariton BEC density far from the excitation spot gradually decreases with the decrease of the 
dielectric constant of the microcavity material, thus semiconductors with higher dielectric constant provide
better conditions for the observation of the polariton BEC propagation.
}

\vspace{-0.5cm}
\section{Conclusion}\label{sec:conclusions}
\vspace{-0.4cm} \setlength{\parskip}{0.1cm plus0mm minus3mm}
Through simulations of the exciton polariton 
BEC  dynamics by using the non equilibrium Gross-Pitaevskii equation,
 we demonstrated that the long-lifetime ($\tau \sim 100$ ps) 
polaritons in a wedge-shaped microcavity can propagate  over a 
macroscopically long distance $\sim 500$ $\mu$m.
This distance is large compared to that for  short-lifetime polaritons 
in traditional experiments \cite{Snoke:02,Kasprzak:06,Amo:09a,Amo:09,Sermage:01}.
The maximum of the polariton density in the BEC is shifted from the center of the excitation spot 
in the direction of the external force due to the exciton-polariton interaction.

We also proposed to observe a polariton BEC propagation in a microcavity with an embedded gapped graphene layer.
It was found that the  BEC density at large distances from the excitation spot 
in a semiconductor quantum well and in gapped graphene
are comparable with each other. However, in graphene there is an additional parameter that controls the 
long-lifetime polariton propagation, which is the  energy of a gap in the electron and hole energetic spectra. 
The obtained results can be useful for practical applications of coherent  polariton flow in a
high-quality microcavity  in working elements of integrated optical circuits \cite{High:08,Liew:10,Menon:10}.
The advantage of graphene in a microcavity  is that the propagation of a polariton BEC 
is dynamically tunable via electrostatic gating.
\vspace{-0.5cm}
\section*{Acknowledgment}\vspace{-0.4cm}
The authors are grateful to D.\ W.\ Snoke for stimulating discussions. 
G.V.K. gratefully acknowledges support from PSC CUNY grant \#65103-00 43. 
{\marko The authors are also grateful to the Center for Theoretical Physics 
of New York City College of Technology of the City University of New York 
for support of the numerical calculations.}

%\bibliography{lowtempgk}
%\bibliography{polaritonprl}

\end{document}